\documentclass[a4paper,UKenglish,lighttt]{oasics}
\usepackage{microtype}
\title{Visualizing the Evaluation of Functional Programs for Debugging}

\author{John Whitington}
\author{Tom Ridge}
\affil{University of Leicester\\
  \texttt{\{jw642, tr61\}@le.ac.uk}}
\authorrunning{J. Whitington and T. Ridge}

\Copyright{John Whitington and Tom Ridge}


\serieslogo{}
\volumeinfo
  {Billy Editor and Bill Editors}
  {2}
  {Conference/workshop/symposium title on which this volume is based on}
  {1}
  {1}
  {1}
\EventShortName{}
\DOI{10.4230/OASIcs.xxx.yyy.p}

\begin{document}

\maketitle

\begin{abstract}In this position paper, we present a prototype of a visualizer for functional programs. Such programs, whose evaluation model is the reduction of an expression to a value through repeated application of rewriting rules, and which tend to make little or no use of mutable state, are amenable to visualization in the same fashion as simple mathematical expressions, with which every schoolchild is familiar. We show how such visualizations may be produced for the strict functional language OCaml, by direct interpretation of the abstract syntax tree and appropriate pretty-printing. We describe (and begin to address) the challenges of presenting such program traces in limited space and of identifying their essential elements, so that our methods will one day be practical for more than toy programs. We consider the problems posed by the parts of modern functional programming which are not purely functional such as mutable state, input/output and  exceptions. We describe initial work on the use of such visualizations to address the problem of program debugging, which is our ultimate aim. 
 \end{abstract}

\section{Introduction}
\label{sec:introduction}

\noindent When we do mathematics on paper, we write an expression or equation and, through a series of legal transformations, produce a simpler one that, we hope, tells us what we want to know. It is the same with functional programming, but the semantics define more closely the order in which the expression is evaluated, choosing each transformation by inspection of the shape of the expression. Of course, this is not quite how the compiled code runs, but it is the mental model. So it is natural, when teaching students functional programming, to proceed by analogy to the mathematical model in which they are already well-practiced. Since computer languages must be more formal in their choice of evaluation order, we tend to underline the sub-expression being evaluated at each step. Given the function \texttt{f} for doubling a number, we might have:
\begin{eqnarray*}
 & & \texttt{\underline{f 3} = 1 + 2 * 3}\\
 \Longrightarrow & & \texttt{6 = 1 + \underline{2 * 3}}\\
 \Longrightarrow & & \texttt{6 = \underline{1 + 6}}\\
 \Longrightarrow & & \texttt{\underline{6 = 7}}\\
 \Longrightarrow & & \texttt{false}
\end{eqnarray*}
There are differences from mathematics, of course: the last step may be rather confusing to the schoolchild (our equals sign does not denote an equation as such, but a comparison operator). Such visualizations are longwinded to write on paper, for all but the least substantial programs. We should like to generate them by computer. In order to provide a tool useful to both learners and the everyday programmer, we will begin with a subset of a real language (rather than building our own toy one), extend it to work for the whole language, and integrate it properly with the toolchain as a first class citizen. We are writing, in essence, a step-by-step interpreter.

What is the relevance to debugging? The dream of debugging is this: having observed a misbehaviour caused by a bug, we assemble all relevant information, both about the program source and the full trace of the program's operation, and, describing ourselves concisely to the computer, we narrow the circumstances down, again and again reducing the search space, until we have the bug in our grasp, and understand it fully. The fix is then often easy. But we are very far from this dream, even today. In his introduction to a Special Issue of Communications of the ACM in 1997 ``The Debugging Scandal and What to Do About It'' \cite{lieberman}, Lieberman  writes: \textit{``Today's commercial programming environments provide debugging tools that are little better than the tools that came with programming environments thirty years ago. It is a sad commentary on the state of the art that many programmers name \textup{``inserting print statements''} as their debugging technique of choice.''} We claim this is still largely true, twenty years later.

\section{Related Work}

\noindent Two useful surveys \cite{UrquizaFuentes}, \cite{survey2009} give a general overview of recent developments in this area, the first specific to functional programming, the second with wider scope. A very broad introduction \cite{PetredeQuincey} gives background. A comprehensive survey \cite{sorva} of education systems for program visualization is useful too. We pick out a few recent systems for further discussion.

The \textit{WinHIPE} system \cite{winhipe} is a recent incarnation of these ideas for the HOPE language. It uses a step-by-step evaluation system, and explicitly addresses the problems of scale by elision of information and a focusing mechanism. The emphasis, however, is on graphical (tree-based) representations, an approach we shall not take, being of the belief that trees can often be, in fact, harder to read than well-pretty-printed program representations. The \textit{Visual Miranda Machine} \cite{visual-miranda} provides a trace of the evaluation of a lazy functional program, together with a commentary showing the reason for choosing each evaluation step. There is a discussion of granularity, taking the example of the ``list comprehension'' language feature. \textit{DrScheme} \cite{drscheme} provides, amongst many other facilities, an ``algebraic stepper'' for the Scheme language that can print out steps of evaluation. The stepper, however, supports only a subset of the language. The implementation is interesting, though -- it reuses some of the underlying Scheme implementation to ensure equivalent semantics.  Touretsky  describes a LISP-based system \cite{touretzky} that produces mainly textual traces, but with some use of graphical elements to indicate the different scoping mechanisms peculiar to LISP. The presentation of \textit{ZStep95} \cite{zstep95} begins by noting that debugging is, essentially, a human interface problem. The authors concentrate on the concept of \textit{immediacy} (temporal, spatial, and so on), which they see as essential, and exhibit a stepping debugger for a functional language which can go back and forth through time. There is a system bolted onto Haskell \cite{Gill00debugginghaskell} which is not an interpreter as such, but allows the programmer to insert observation points in the code, within the syntax of the source language. At such points, the values are printed out. This has the advantage of explicitly dealing with at least part of the problem of scale -- only what we wish to be printed is printed.  Another approach to this problem is as a special case of the more general concept of a \textit{calculator} \cite{Reeves95thecalculator}, \cite{miracalc}, showing how it pertains to various logical systems with a mathematical basis, not just functional programs. \textit{Prospero} \cite{taylor-thesis} is a more fully-developed system, again for a lazy language. It includes methods for filtering the evaluation trace to elide information and a careful discussion of usability issues.

These systems are mostly concerned with program visualization for teaching; we wish to bias ourselves towards the task of general debugging, hoping that some of the educational uses will be subsumed by it (the authors of DrScheme \cite{drscheme} urge caution here, choosing instead to build a ``tower'' of syntactically restrictive variants of Scheme specifically for educational purposes. They say that, due to the fact that so many sequences of characters are syntactically valid in Scheme, error messages are less confusing when the dialect is restricted -- we would prefer to avoid this in the name of universality).

It is worth pointing out that much research in software visualization concerns overtly graphical approaches. We take a simpler line, sticking to pretty-printing. We claim that the most important aspect of a successful visualization is elision -- reducing the information visible to just what is required so that large datasets may be understood easily -- whether interactively or not. Programmers are used to seeing their program as text, and visualizing its evaluation as, for example, a graphical tree structure, is less useful for debugging large programs (it can be useful, of course, for visualizing program source code structure as opposed to evaluation traces).

\section{Rationale}

\noindent We surveyed users of the strict functional programming language OCaml informally to ask whether they routinely used debuggers, and if not, why not. The overwhelming result was that debuggers are not widely used. The Haskell community has found the same \cite{marlow}. Several respondents honed in on a theme: \textit{``I use tools that I am familiar with when debugging because I don't want to focus on two things (learning a new tool and tracking down/fixing a bug).''} One coined this the ``lack-of-use vicious circle'': \textit{``When you really need a debugger, you're not willing to learn a new tool. When you're willing to learn a new tool, you don't really want to learn a debugger.''}

Wadler records the story \cite{Wadler} of the Standard ML debugger \cite{tolmach} that was deeply intertwined with the compiler and runtime of the SML/NJ Standard ML compiler. As the SML/NJ implementation evolved, the debugger fell out of step, and is no longer available.

The implementations described above all suffer, to a lesser or greater
extent, from a lack of what Marlow\cite{marlow} calls  
\textit{accessibility}. They provide only for a subset of the language, or
require changes to be made to build environments, or do not scale well. So there is often one or more fundamental impediments to their use -- they are not \textit{accessible}. A debugger must be as accessible as a compiler. Marlow claims that the most complete Haskell trace debugger, Hat \cite{hat}, remains largely unused due to a lack of such accessibility -- for example, it must be modified to support new third-party libraries. We
intend, then, to bake in the correct design decisions to support widespread
applicability (and thus adoption) from the beginning, even if it is at the
expense of other desirable characteristics (such as speed). We aim for our system to a) be able
to support the whole language by design; b) be suitable for any build
environment where OCaml programs can already be built; and c) be abstracted from
the compiler, and thus be robust to advances in the language and runtime
environment. Thus, instead of imagining the perfect visualization, writing a toy system, and worrying about how to extend it to a practical one later, we will make design decisions based on the practicalities, and work backward from our goal. Even if our system is initially a toy in the sense that it does not support the full language, it is not a toy in terms of its integration with the language and runtime, and so extending it to the full language should be technologically straightforward (though a sizeable piece of work). Our litmus tests are these: 1) Can our system be used to debug any OCaml program where source is available, even if uses external libraries? 2) Can our system support development of a complex system such as the OCaml compiler itself? 3) Most importantly, of course, do people actually choose to use it?

We shall, therefore, take an extremist approach: we shall worry about Marlow's \textit{accessibility} foremost, and everything else second. Wadler \cite{Wadler} writes: \textit{``\ldots there are few debuggers or profilers for strict [functional] languages, perhaps because constructing them is not considered research. This is a shame, since such tools are sorely needed, and there remains much of interest to learn about their construction and use.''} We aim to right this wrong.

\section{Simple Visualizations}

\noindent As is traditional, we consider a program for calculating the factorial of a positive number, with \texttt{4} as our input:

\medskip
{\small\ttfamily
\noindent\texttt{\textbf{let rec} factorial n =}\\
\texttt{\-~~\textbf{if} n = 1 \textbf{then} 1 \textbf{else} n * factorial (n - 1)}\\
\texttt{\textbf{in}}\\
\texttt{\-~~factorial 4}
}
\medskip

\noindent The upper portion of Figure 1 shows a naive evaluation of this program. This is certainly  not how we would write such an evaluation on paper. Although the evaluation shown is self-contained in the sense that each line of it is a valid program, which might seem a useful property, it is hard to see what is going on. It is large, both in width (how long the expression becomes) and length (how many lines are needed). Writing each evaluation step over multiple lines as we did with the original program above would not only increase the length, but make it difficult to visually compare adjacent lines. We must reduce the amount of information shown, even in this simple case.

{Look now at the lower part of Figure 1, which shows the output of our prototype system. The following differences are apparent: a) We have removed the definition of the factorial function itself. Since it is recursive, its name will appear in the expression anyway; b) We have avoided printing any reduction step which leads to an expression such as \texttt{\textbf{if}\! false} or \texttt{\textbf{if}\! true}; c) We have not shown the intermediate steps of simple arithmetic which reduce \texttt{4\!~*\! (3\! *\! (2\! *\! 1))} to \texttt{24}; d) We have removed trivial arithmetic (like subtracting one), even when it involves variable names, such as reducing \texttt{n - 1} to \texttt{3} directly rather than via \texttt{4\! -\! 1}; e) We have removed \texttt{\textbf{let}} bindings which apply to the whole expression to the left hand side of the \texttt{=>} arrow to avoid too many \texttt{\textbf{let}\! n\! =\! \ldots} instances making the expression too wide; f) We have used some simple syntax colouring in the form of bold for keywords; and g) We have underlined the expression to be reduced at each step. All these changes have been made automatically. Each step is no longer a valid OCaml program, but the increase in readability is significant. Clearly, for larger programs, such elision will be even more important, since the focus needs to be on the currently-evaluating subexpression of a potentially huge expression representing the whole program. Note that all the intervening steps of the computation are performed, but certain lines are not printed. This means that the finer details of the computation may be inspected upon demand.\par}

In the program trace we have already exhibited, it is clear that for realistic programs, the program trace (both its width and its length) may be significant. This issue is discussed in some detail by Taylor \cite{taylor-thesis} and Pajera-Flores \cite{winhipe}. A practical solution, we claim, must involve providing ways of a) eliding information within a single step (reducing the width); b) eliding whole steps (reducing the length); c) searching the resultant trace, if it is still too large to spot the bug; and d) moving backward and forward through the trace to connect cause and effect in the computation.

\begin{sidewaysfigure}
\scalebox{0.65}{
\begin{minipage}{2\textwidth}
{\ttfamily
~~~~let rec factorial n = if n = 1 then 1 else n * factorial (n - 1) in factorial 4\\
=>  ~let rec factorial n = if n = 1 then 1 else n * factorial (n - 1) in let n = 4 in if n = 1 then 1 else n * factorial (n - 1)\\
=>  ~let rec factorial n = if n = 1 then 1 else n * factorial (n - 1) in let n = 4 in if false then 1 else n * factorial (n - 1)\\
=>  ~let rec factorial n = if n = 1 then 1 else n * factorial (n - 1) in let n = 4 in n * factorial (n - 1)\\
=>  ~let rec factorial n = if n = 1 then 1 else n * factorial (n - 1) in let n = 4 in 4 * factorial (n - 1)\\
=>  ~let rec factorial n = if n = 1 then 1 else n * factorial (n - 1) in 4 * factorial (4 - 1)\\
=>  ~let rec factorial n = if n = 1 then 1 else n * factorial (n - 1) in 4 * factorial 3\\
=>  ~let rec factorial n = if n = 1 then 1 else n * factorial (n - 1) in 4 * (let n = 3 in if n = 1 then 1 else n * factorial (n - 1))\\
=>  ~let rec factorial n = if n = 1 then 1 else n * factorial (n - 1) in 4 * (let n = 3 in if false then 1 else n * factorial (n - 1))\\
=>  ~let rec factorial n = if n = 1 then 1 else n * factorial (n - 1) in 4 * (let n = 3 in n * factorial (n - 1))\\
=>  ~let rec factorial n = if n = 1 then 1 else n * factorial (n - 1) in 4 * (let n = 3 in 3 * factorial (n - 1))\\
=>  ~let rec factorial n = if n = 1 then 1 else n * factorial (n - 1) in 4 * (3 * factorial (3 - 1))\\
=>  ~let rec factorial n = if n = 1 then 1 else n * factorial (n - 1) in 4 * (3 * factorial 2)\\
=>  ~let rec factorial n = if n = 1 then 1 else n * factorial (n - 1) in 4 * (3 * (let n = 2 in if n = 1 then 1 else n * factorial (n - 1)))\\
=>  ~let rec factorial n = if n = 1 then 1 else n * factorial (n - 1) in 4 * (3 * (let n = 2 in if false then 1 else n * factorial (n - 1)))\\
=>  ~let rec factorial n = if n = 1 then 1 else n * factorial (n - 1) in 4 * (3 * (let n = 2 in n * factorial (n - 1)))\\
=>  ~let rec factorial n = if n = 1 then 1 else n * factorial (n - 1) in 4 * (3 * (let n = 2 in 2 * factorial (n - 1)))\\
=>  ~let rec factorial n = if n = 1 then 1 else n * factorial (n - 1) in 4 * (3 * (2 * factorial (2 - 1)))\\
=>  ~let rec factorial n = if n = 1 then 1 else n * factorial (n - 1) in 4 * (3 * (2 * factorial 1))\\
=>  ~let rec factorial n = if n = 1 then 1 else n * factorial (n - 1) in 4 * (3 * (2 * (let n = 1 in if n = 1 then 1 else n * factorial (n - 1))))\\
=>  ~let rec factorial n = if n = 1 then 1 else n * factorial (n - 1) in 4 * (3 * (2 * (let n = 1 in if true then 1 else n * factorial (n - 1))))\\
=>  ~4 * (3 * (2 * 1))\\
=>  ~4 * (3 * 2)\\
=>  ~4 * 6\\
=>  ~24\par}
\bigskip

{\ttfamily~~~~factorial 4\\
n = 4 ~=> ~\underline{\textbf{if} n = 1 \textbf{then} 1 \textbf{else} n * factorial (n - 1)}\\
n = 4 ~=> ~n * factorial (\underline{n - 1})\\
\-~~~~~~~=> ~4 * \underline{factorial 3}\\
n = 3 ~=> ~4 * (\underline{\textbf{if} n = 1 \textbf{then} 1 \textbf{else} n * factorial (n - 1)})\\
n = 3 ~=> ~4 * (n * factorial (\underline{n - 1}))\\
\-~~~~~~~=> ~4 * (3 * \underline{factorial 2})\\
n = 2 ~=> ~4 * (3 * (\underline{\textbf{if} n = 1 \textbf{then} 1 \textbf{else} n * factorial (n - 1)}))\\
n = 2 ~=> ~4 * (3 * (n * factorial (\underline{n - 1})))\\
\-~~~~~~~=> ~4 * (3 * (2 * \underline{factorial 1}))\\
n = 1 ~=> ~4 * (3 * (2 * (\underline{\textbf{if} n = 1 \textbf{then} 1 \textbf{else} n * factorial (n - 1)})))\\
\-~~~~~~~=> ~4 * (3 * (\underline{2 * 1}))\\
\-~~~~~~~=>* 24\par}
\end{minipage}}
\bigskip

\caption{A naive rendering of the evaluation of \texttt{factorial\! 4} showing each step of the evaluation, followed by an automatically trimmed one, eliding a) parts of the evaluation of the \texttt{\textbf{if}} construct; b) the definition of a recursive function mentioned in the expression; c) the final portion of arithmetic and d) trivial operations such as \texttt{3\! -\! 1}. In addition, \texttt{\textbf{let}} expressions unique in the whole expression are written on the left, and simple syntax highlighting has been used. The expression to be reduced in each step is underlined.}\end{sidewaysfigure}

If we wish to be able to reduce traces on command to find a bug in a morass of data, we shall need a concise and powerful way for the programmer to describe the elisions and searches required. That is to say, we need to have a way for the programmer to translate \textit{``I'm sure this bug has something to do with the} \texttt{tree} \textit{data type, and I know it must happen after the tree has been populated, but before the last element is removed. I know the proximate cause is a} \texttt{Not\_found} \textit{exception being raised.''} into something the computer can understand, and which results in a reduced, concise, useful trace.

\section{Other Kinds of Computation}
\noindent Although the primary challenge is one of scale, we must consider also non-functional computation (such as the use of input/output or mutable data). There follows a brief discussion of several of these, to give a flavour of the complications involved. There are plenty of others, of course. For example, we have yet to explore the visualization of concurrent or parallel execution.

\subsection{Exceptions}
Though exceptions can be explained using the same term-rewriting rules as any other functional construct, it is likely that some special treatment will be necessary, especially for the visualization of larger programs. Some of the complications of exception visualization for imperative languages are described by Shah \cite{shah}, many of which will apply in our case too. Exceptions are important, of course, because they are used for dealing with illegal states, a common cause of problems which we end up debugging. Additionally, and somewhat unusually, exceptions are frequently used in OCaml programs not just for genuinely exceptional situations, but for local control flow. So it is especially important that the debugger's approach to exceptions is lightweight.

\subsection{Input/Output and System Primitives}

Consider the following program, which reads a line from standard input, and then prints it on standard output:
{\small\begin{verbatim}
print_string (input_line stdin)
\end{verbatim}}
\noindent Do we separate the output of the program from the output of the debugger, or is it better to show everything interleaved? In addition, how do we deal with \texttt{print\_int} and \texttt{input\_line}, which are members of the most basic parts of the Standard Library, themselves defined in terms of system primitives? In the present prototype, the output of the program and the debugger are interleaved. Here is a session, with the Standard Library and primitive operations elided, which is the default:

{\small\begin{verbatim}
    print_string (input_line <in_channel>)
SLATE
SLATE=>  ()
\end{verbatim}}

\noindent (We typed the word ``SLATE''). If the user really wants the gory details of the inside of the Standard Library, they can be shown instead:

\medskip

{\small\ttfamily
\noindent~~~~print\_string (input\_line <in\_channel>)\\
=>~~print\_string (\textbf{let} x = <in\_channel> \textbf{in} <<input\_line>>)\\
=>~~print\_string <<input\_line>>\\
SLATE\\
=>~~print\_string "SLATE"\\
=>~~\textbf{let} x = "SLATE" \textbf{in} output\_string <out\_channel> x\\
=>~~\textbf{let} x = "SLATE" \textbf{in} (\textbf{let} x = <out\_channel> \textbf{in} \textbf{fun} y -> <<output\_string>>) x\\
=>~~\textbf{let} x = "SLATE" \textbf{in} (\textbf{fun} y -> \textbf{let} x = <out\_channel> \textbf{in} <<output\_string>>) x\\
=>~~(\textbf{fun} y -> \textbf{let} x = <out\_channel> \textbf{in} <<output\_string>>) "SLATE"\\
=>~~\textbf{let} y = "SLATE" \textbf{in} \textbf{let} x = <out\_channel> \textbf{in} <<output\_string>>\\
=>~~\textbf{let} y = "SLATE" \textbf{in} <<output\_string>>\\
=>~~<<output\_string>>\\
SLATE=>~~()
\par}

\medskip

\noindent Abstract data types for standard input and output channels have been written \texttt{<thus>}. The invocation of a genuine system primitive (rather than just a standard library function) is written with double angle brackets around it \texttt{<<thus>>}. You can see why it is usually sensible to elide such information. We are not typically debugging libraries, but our own program. Until we are sure a library is at fault, we don't want to delve into its internals. This is just one example of the importance of mechanisms of elision in trace debugging. 

\subsection{Mutability}

The OCaml language is what has been termed a ``functional-first'' language. That is to say, whilst we may tend to write in a pure functional style, we may also use mutable cells to hold changing values. Thus, it is necessary either to ensure that the contents of a cell is always displayed in the printed evaluation step, or that there is another way for the user to see it or request to see it. In the latter case, which is probably preferable, we may simply use techniques from the broad range available in traditional debuggers for imperative languages.

\section{Implementation Notes}

\noindent The present prototype provides visualization for a subset of the OCaml language, together with a number of methods for elision of information to produce more reasonable traces. Our interpreter is, thus far, only a thousand lines of code, due to the ease with which we can use parts of the OCaml compiler: in recent versions of OCaml the compiler is built not only in executable form, but in library form as \textsf{compiler-libs}. This means that one may write a program which uses types and functions from the OCaml compiler, for example the Abstract Syntax Tree type. Our \texttt{ocamli} interpreter is a program of this form: it uses the standard lexer, parser and type-checker direct from \textsf{compiler-libs}. And so, writing an interpreter for OCaml programs has required almost no duplication or modification of compiler code. This has reduced vastly the cognitive load of such an endeavour, and increased the likelihood that the interpreter will, with little modification, continue to build and function well into the future. Before \textsf{compiler-libs}, we would have had to copy reams of code from the OCaml compiler source code, or provide our interpreter as a patch to the OCaml compiler itself. Both are inadvisable from the  perspective of Marlow's accessibility -- that a debugger should be available in a low-friction manner to users, for any project, at any time.

The prototype implementation is simplistic, using a number of tree-processing passes to perform the evaluation and elisions, so does not guarantee that the time or space complexity of interpreting a program is the same as the time or space complexity of compiling and running that program. We hope to discover in future to what extent such a guarantee is possible (at least modulo the pretty-printing -- clearly, printing out all the stages of a computation cannot help but increase the time complexity of interpreting it).

\section{Conclusion}
\noindent We have advanced an ambitious but, in our opinion, practical scheme for the implementation of an interpretive debugger for a popular functional language, and described aspects of the current prototype. There is much to be done. Will it fall by the wayside like so many other debuggers? Or have we found the right formula?

\bibliography{visfunc}

\end{document}